\documentclass[fleqn,10pt]{SelfArx} 

\definecolor{color1}{RGB}{0,0,90} 
\definecolor{color2}{RGB}{0,20,20} 

\usepackage{hyperref} 
\hypersetup{hidelinks,colorlinks,breaklinks=true,urlcolor=color2,citecolor=color1,linkcolor=color1,bookmarksopen=false,pdftitle={Title},pdfauthor={Author},urlcolor=blue}

\usepackage{graphicx}
\usepackage[square]{natbib}
\usepackage{amsmath}
\usepackage{multirow}
\usepackage{subfigure}
\usepackage{multirow}

\newcommand{\n}[1]{\mathrm{#1}}

\JournalInfo{Published in Energy Conversion and Management, Vol. 119 , 473-480, 2016} 
\Archive{\href{http://dx.doi.org/10.1016/j.enconman.2016.04.042}{DOI: 10.1016/j.enconman.2016.04.042}} 

\PaperTitle{A thermoelectric power generating heat exchanger: Part I - Experimental realization} 

\Authors{R. Bj\o{}rk\textsuperscript{1}, A. Sarhadi\textsuperscript{1}, N. Pryds\textsuperscript{1}, N. Lindeburg\textsuperscript{2} and P. Viereck\textsuperscript{2}} 
\affiliation{\textsuperscript{1}\textit{Department of Energy Conversion and Storage, Technical University of Denmark - DTU, Frederiksborgvej 399, DK-4000 Roskilde, Denmark}} 
\affiliation{\textsuperscript{2}\textit{Peltpower ApS, Storkev\ae{}nget 14, DK-4653 Karise, Denmark}} 
\affiliation{*\textbf{Corresponding author}: rabj@dtu.dk} 

\newcommand{\keywordname}{Keywords} 

\Abstract{An experimental realization of a heat exchanger with commercial thermoelectric generators (TEGs) is presented. The power producing capabilities as a function of flow rate and temperature span are characterized for two different commercial heat transfer fluids and for three different thermal interface materials. The device is shown to produce 2 W per TEG or 0.22 W cm$^{-2}$ at a fluid temperature difference of 175 $^\circ$C and a flow rate per fluid channel of 5 L min$^{-1}$. One experimentally realized design produced 200 W in total from 100 TEGs. For the design considered here, the power production is shown to depend more critically on the fluid temperature span than on the fluid flow rate. Finally, the temperature span across the TEG is shown to be 55\% to 75\% of the temperature span between the hot and cold fluids.}

\begin{document}

\flushbottom 

\maketitle 


\thispagestyle{empty} 

\section{Introduction}
Thermoelectric generators (TEG) are solid-state physical devices that convert part of a heat flux directly into electricity using the Seebeck effect. Using TEGs to partially convert waste heat directly into electricity has long been considered an ideal use of TEGs.

Much focus in recent years has been on the hot waste exhaust gas from vehicle engines \citep{Yang_2009,Salzgeber_2010}, but here we will consider a more general style of power generating heat exchanger, which can be adapted to many different kinds of applications with waste heat. The generic setup that we consider is a heat exchanger with two different fluids, hot and cold respectively, and TEGs placed in between these, such that they produce electricity due to the temperature difference and subsequent flow of heat. Such a universal device with heat transfer fluids have been considered from a theoretical perspective in literature for at least 35 years \cite{Henderson_1979}. The novelty of such a device is that the device itself does not need to be adapted to the heating or cooling sources, but relays only on a suitable flow of hot and cold fluids, and the device can thus be adapted to a large range of waste heat applications.

At present, several different directions are considered in the design of a TEG-equipped heat exchanger. One direction is a TEG micro-heat exchanger, which has been considered by \citet{Rezania_2013}, \citet{Wojtas_2013} and \citet{Barry_2015}. Another direction of design considered is heat exchangers using air as one of the heat transfer fluids \cite{Belanger_2012}, where a power generating heat exchanger was shown to produce a power of 110 W from 279 TEG modules from a domestic atmospheric gas boiler \cite{Martinez_2010}. In a somewhat conceptually similar system, \citet{Zhang_2015} considered an air to air heat exchanger with TEGs and a finned heat exchanger design. Using 112 TEG, a total power of 250 W was produced using the exhaust from a diesel generator. Finally, \citet{Favarel_2016} examined a TEG heat exchanger using air as the hot heat transfer fluid, and showed a good agreement ($\le$ 10\%) between an experimental system and a numerical model. The system produced between 15-50 W. Another popular topic is TEG equipped heat exchangers for automotive applications. Here \citet{Lu_2015} used a numerical model to conclude that a metal foam heat exchanger for the exhaust gas can produce a larger total power output and efficiency for a TEG heat exchanger compared to a rectangular offset-strip fin. The more conventional heat exchanger geometry was examined by \citet{Tao_2015}, while \citet{Su_2015} considered the same application but with a folded-shape internal structure heat exchanger. The final design direction is that of TEG fluid heat exchangers, where a number of experimental realizations exist. \citet{Lu_2012} considered integrating thermoelectric devices into a large-capacity heat exchanger (100 kW), and experimentally realized a small prototype. However, the goal of the device was to provide supplemental cooling and not to generate power, as is considered here. This was also the case of the system studied by \citet{Campbell_2013}. Concerning an actual power producing device, \citet{Gou_2010} demonstrated an experimental TEG-parallel plate heat exchanger with hot water on one side and a flow of cold air on the other side, using ten TEGs and a hot flow rate of 6 L min$^{-1}$. This device realized a temperature difference of 18 K and generated a power of 0.9 W, with a numerical model indicating more power could be generated at higher temperature differences. \citet{Esarte_2001} considers numerically and experimentally a TEG-heat exchanger with a spiral, zig-zag or straight fins and with water as the heat transfer fluid. \citet{Suzuki_2012} used a numerical model of a simple TEG-parallel plate heat exchanger to consider how to best introduce the fluid flows across the TEGs. \citet{Niu_2009} built an experimental 56 TEGs heat exchanger using copper plates, a 60:40 glycol/water mixture and commercial bismuth telluride TEGs. The heat exchanger was so efficient that the temperature span across the TEGs equalled the fluid inlet temperatures, leading to maximally efficient TEGs. The total power produced was above 160 W at a flow rate of 6.6 L min$^{-1}$. \citet{Crane_2004} built a simple TEG-heat exchanger with hot water/glycol and cold air as heat transfer fluids and found an optimum power per cost around 1.1 kW/\$10000. In another setup \citet{Yu_2007} used a numerical study of a parallel plate co- and counter flow heat exchanger with TEGs to shown that at an inlet temperature of 200 $^\circ$C the total power of 1.8 W per TEG could be produced for a flow rate of 4 L min$^{-1}$ of pressurized water. Finally, \citet{Champier_2010} constructed an experimental biomass cook stoves equipped with 4 TEGs and showed that this could produce close to 8 W at a temperature difference of 170 K.

The conclusion of the existing investigations on TEG-fluid heat exchanger systems remains broad. It has been shown that power generating devices can be realized, but an understanding of how to realize a system with a high temperature span across the TEGs is still challenging. This temperature span is influenced primarily by geometry, flow rate and thermal interfaces, although secondary factors can also be present. In this paper we consider an experimental TEG-fluid heat exchanger and analyze the performance as a function of heat transfer fluid and thermal interface material, as well as flow rate and temperature difference across the TEGs. This will give an understanding of which parameters affect power generation, as well as help in understanding how to realize a cost-effective system.

Previous investigations have not consider the influence of the individual parameters on the system performance, but have focussed more on constructing a working system, capable of generating power. By elucidating the influence of the system parameters, an in-depth understand of how to build the most efficient power generating heat exchanger can be achieved.

\section{The experimental setup}
\begin{figure}[!t]
  \centering
  \includegraphics[width=1\columnwidth]{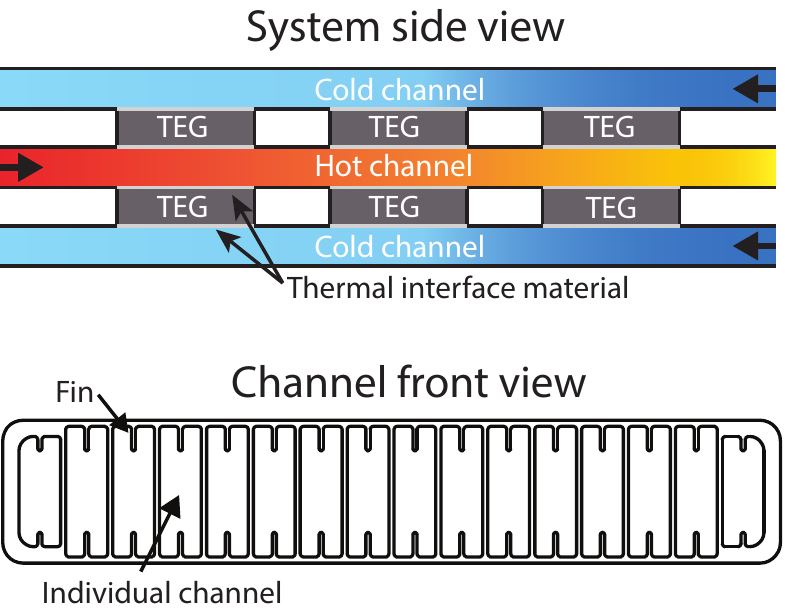}
  \caption{A schematic illustration of the experimental setup. A number of TEGs, here six, are sandwiched in between one hot and two cold flow channels. A thermal interface material in placed on either side of the TEGs. In the space between the TEGs, an insulating wool was placed. The specific geometry of the flow channels used are also shown. The dimensions of these are given in Table \ref{Table.Pipe_measurements}.}
  \label{Fig_Schematic}
\end{figure}

\begin{figure}[!t]
  \centering
  \includegraphics[width=1\columnwidth]{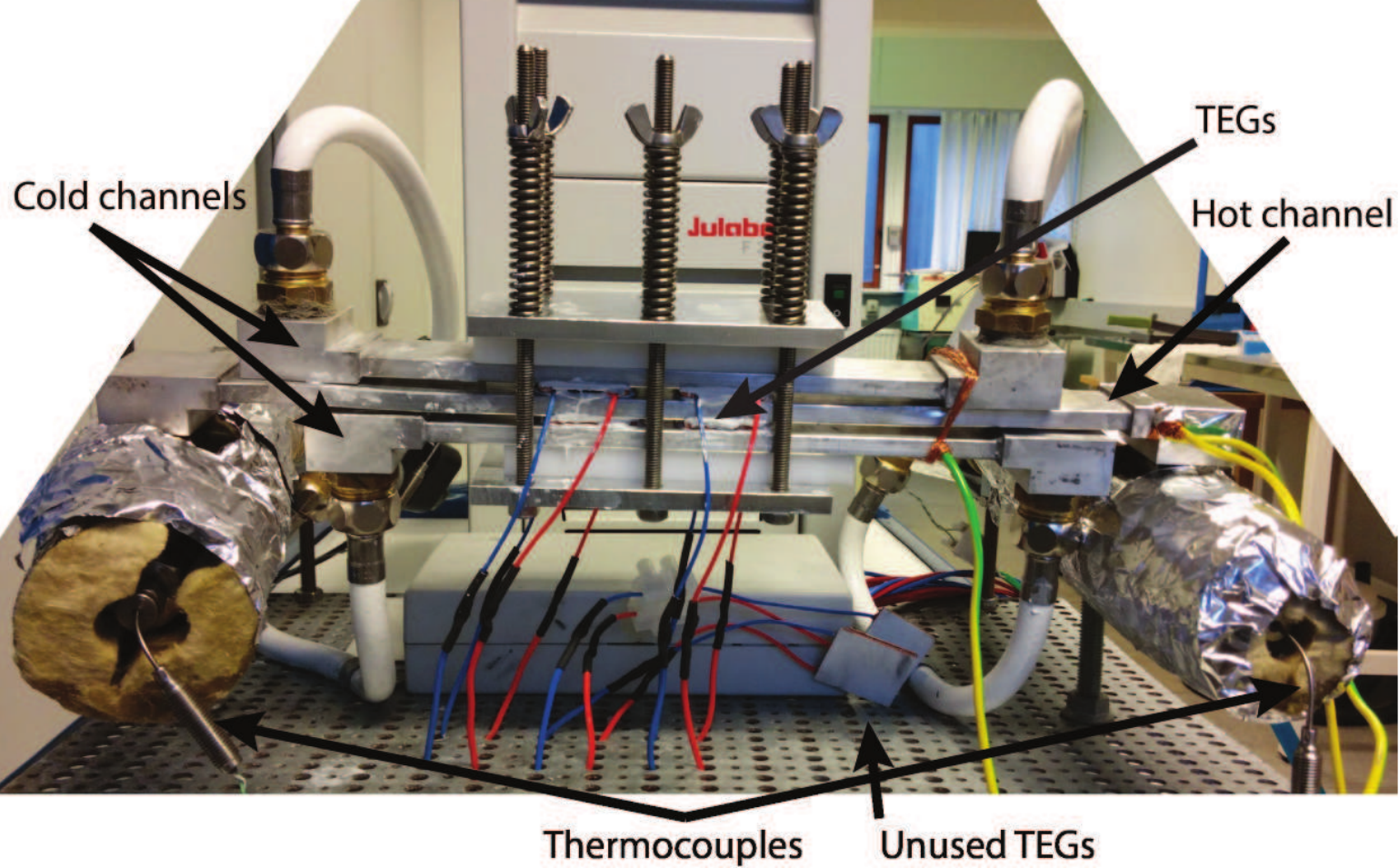}
  \caption{A front side image of the experimental setup, with the various components indicated. In the pictured experiment only four of the available TEGs were used.}
  \label{Fig_Img_for_article}
\end{figure}

Initially, an small scale experimental realization of a TEG heat exchanger was built and tested. The device consisted of a number of TEGs (4 or 6), with performance characteristics described below, sandwiched between aluminium fluid channels containing a hot and cold heat transfer fluid, respectively. A schematic of the system is shown in Fig. \ref{Fig_Schematic}, along with a drawing of the specific pipes used for the fluid channels (PTT001 from Sapa Precision Tubing, T\o{}nder, Denmark). The measured dimensions of the pipe, as determined using an optical microscope, are given in Table \ref{Table.Pipe_measurements}. An image of the actual experimental setup can be seen in Fig. \ref{Fig_Img_for_article}. The flow system is run in counterflow mode, i.e. with opposite directions of the fluid flow in the hot and cold channels. This is done to ensure the same temperature span across TEGs all along the channel. In co-flow setup, the TEG at the channel inlet would experience a significantly larger fluid temperature span than the TEG at the channel outlet. The specific channels were selected based on availability, in order to reduce cost. However, previous studies have shown that the exact geometry of the fins themselves seems to have a relative small influence on the performance of the system, albeit for system where the heat transfer fluid was air \citep{Belanger_2012}, or the channels were micro-sized \cite{Wojtas_2013}. As can also be seen in Fig. \ref{Fig_Img_for_article}, the channels and TEGs are spring loaded, thus allowing the pressure on the system to be controlled.

\begin{table}[!t]
\begin{center}
\caption{The measured dimensions of the components of the pipe.}\label{Table.Pipe_measurements}
\begin {tabular}{lr}
Component & Value\\ \hline
Width of pipe      & 38 mm\\
Height of pipe: & 7 mm \\
Width of pipe wall & 320 $\mu$m \\
Width of individual channels & 2000 $\mu$m \\
Height of fin & 950 $\mu$m \\
Width of fin wall & 220 $\mu$m
\end {tabular}
\end{center}
\end{table}

The flow rate of the hot fluid was measured using a Hedland H602A-002-HT high temperature flow meter, which was calibrated individually for each heat transfer fluid. The temperature of the hot fluid was controlled using a Julabo F25 heating circulator with built-in pump. The flow was regulated using a valve. For all experiments, the cold heat transfer fluid was water at 25 $^\circ$C, controlled using a chiller. The flow rate for the cold water was measured using an electronic flow meter. The flow rate was kept between 6.3-7.9 L min$^{-1}$ for all experiments, as this was sufficient to keep the cold side temperature stable. The pressure drop was not measured across the system, but as square channels with simple fins were used, this can be estimated analytically. The pressure drop, $\Delta{}p$, is calculated using the Hagen-Poiseuille equation,
\begin{equation}
  \Delta{}p = \frac{128\mu{}L\dot{V}}{\pi{}D_\n{H}^{4}}
\end{equation}
where $\mu$ is the dynamic viscosity, $L$ is the length of the pipe, $\dot{V}$ is the volumetric flow rate and $D_\n{H}$ is the hydralic diameter, here $D_\n{H} = 2wh/(w+h)$ where $w$ is the width and $h$ is the height of the channel. Using the width and height of the channels given in Table \ref{Table.Pipe_measurements} and tabulated values of the viscosity, the total pressure drop in all channels can be calculated. The theoretical pumping work is the sum of the pumping work of the hot side fluid and the cold side fluid, where the latter is always water. Two hot side fluids were considered, as will be described subsequently. For both fluids the maximum total theoretical pumping work is close to 0.25 W per TEG, which occur at the highest flow rate and lowest temperature of the hot side fluid. However, at high temperature above 150 $^\circ$C for both fluids the pumping work decreases to less than 0.025 W per TEG due to a decrease in viscosity with temperature.

\subsection{Experimental uncertainties}
In this paragraph the experimental uncertainty associated with the used measurement equipment will be described. Regarding the temperature measurements, the temperature of the fluid is measured using type K thermocouples connected to a Omega TC-08 USB thermocouple data acquisition module. The accuracy of the type K thermocouple is $\pm$2.2 $^\circ$C in the temperature range investigated here, while the TC-08 has an accuracy of the sum of $\pm$0.2\% of the reading and $\pm$0.5 $^\circ$C. At the highest fluid temperature range, the absolute temperature accuracy is 2.3 $^\circ$C. However, the resolution is less than 0.025 $^\circ$C. The temperature stability of the Julabo F25 chiller is $\pm$0.01 $^\circ$C.

The accuracy of the Hedland H602A-002-HT flow meter is $\pm$ 2\%, i.e. 0.15 L min$^{-1}$. This is in line with the fact that the flow meter has to be read by eye, with indicator marks for every 0.25 L min$^{-1}$. The repeatability of flow measurements is given as $\pm$1\%.

The electrical measurements were done using a Keithley 2700 Digital MultiMeter with a 20 channel Keithley 7700 board. For the voltage measurements, the resolution is 10$\mu$V and the accuracy is 0.0030\%, while for the current measurements the resolution is 1$\mu$A and the accuracy is 0.06\%.

The largest experimental uncertainties are thus related to the measurements of the flow rate. However, these are still acceptable for the conducted experiments, where the flow rate was typically varied in steps larger than 1.5 L min$^{-1}$, as will be described subsequently.

\subsection{Characterization of the TEGs}
The TEGs used in the experimental setup were commercial modules of type TG12-4 from Marlow Industries, Inc. These are made of $n$- and $p$-type bismuth telluride, Bi$_2$Te$_3$ (BiTe), which is the thermoelectric material with the highest efficiency at hot side temperatures below 250 $^\circ$C. The area of a TEG is 30x30 mm$^2$. A number of other commercial Bi$_2$Te$_3$ modules exist, with performances more or less identical to the modules used here.

\begin{figure*}[!t]
\centering
\subfigure[Efficiency]{\includegraphics[width=1\columnwidth]{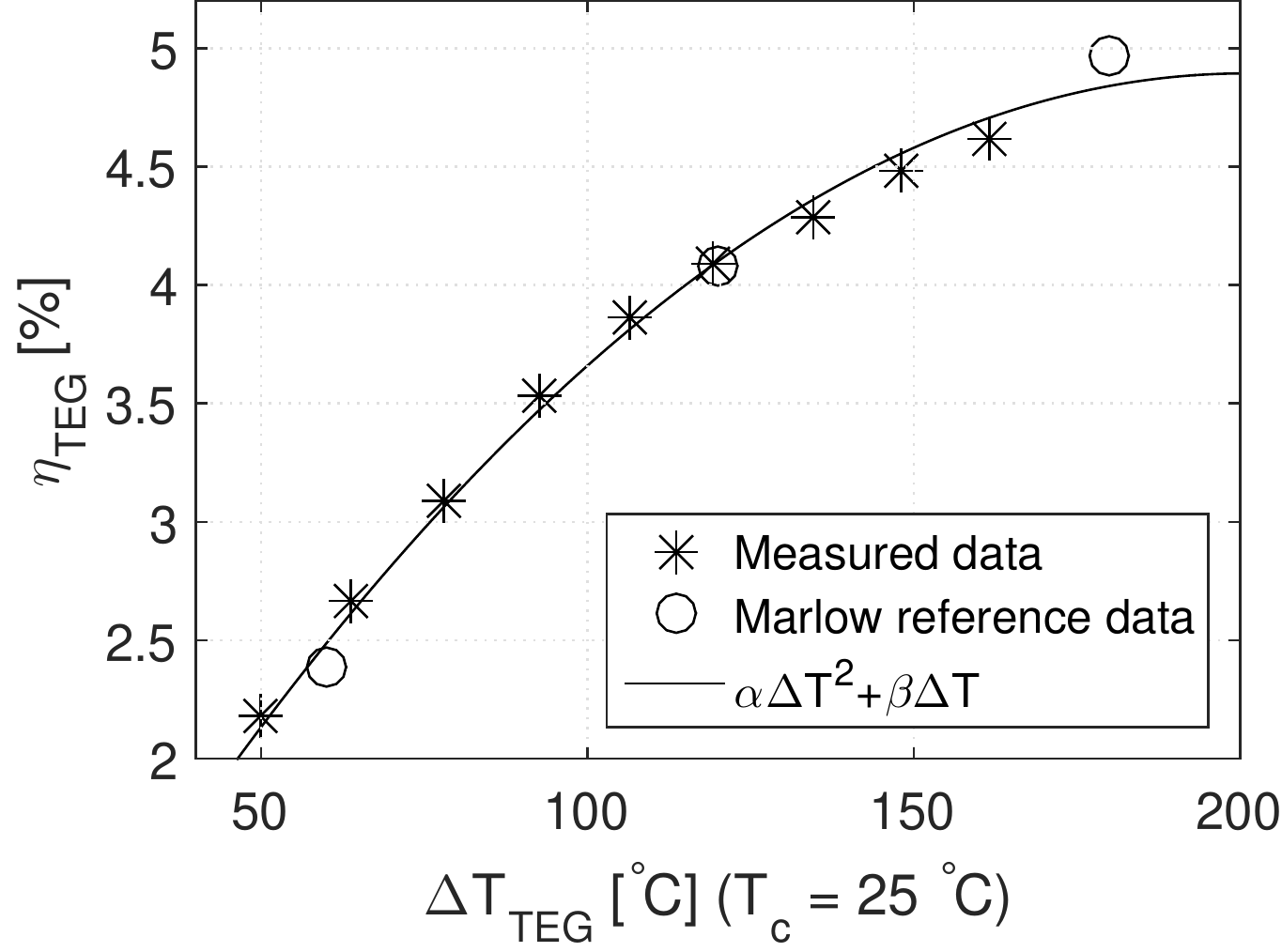}}\hspace{0.2cm}
\subfigure[Temperature span]{\includegraphics[width=1\columnwidth]{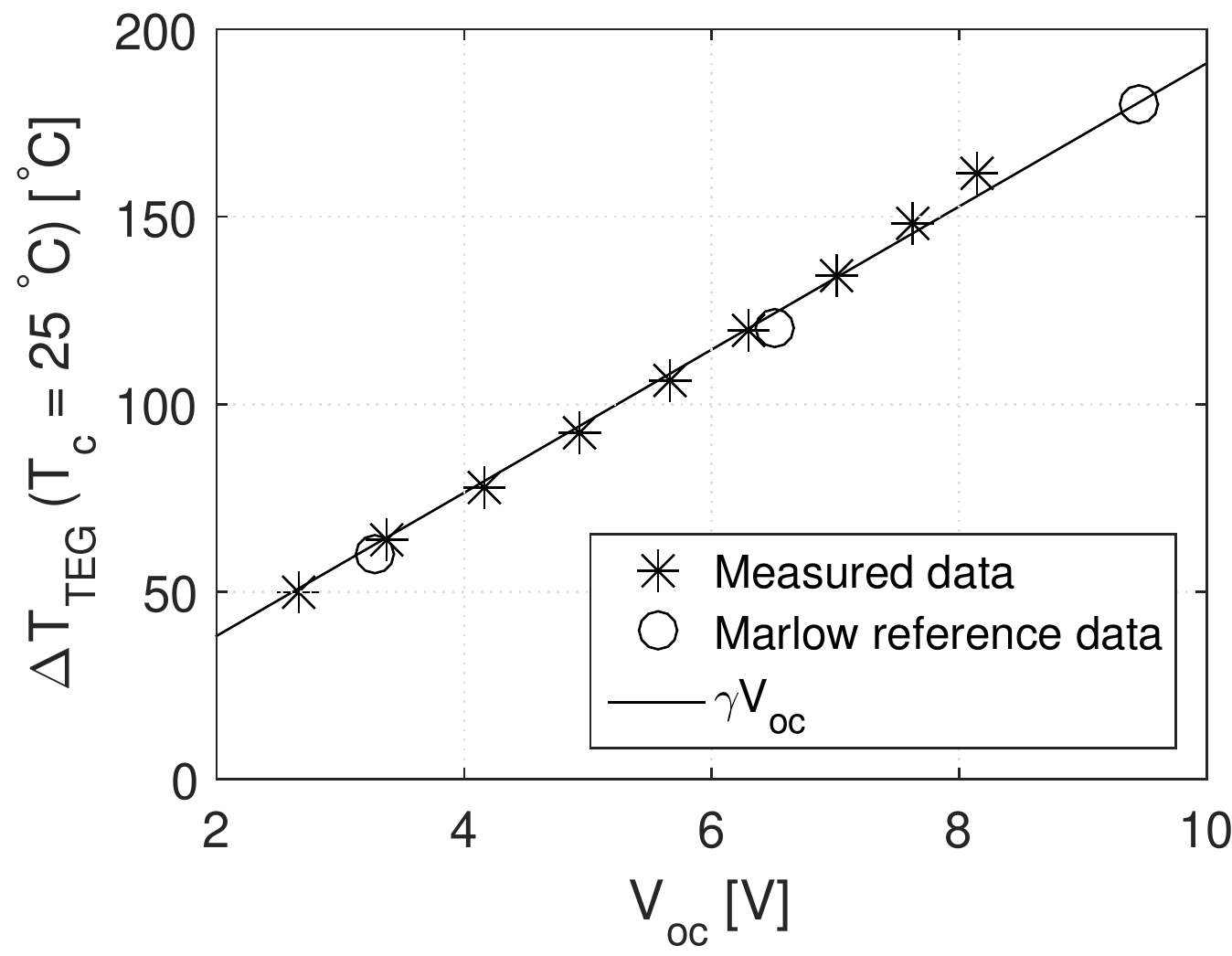}}\hspace{0.2cm}
\caption{(a) The measured efficiency as a function of temperature span (a) and temperature span as a function of open circuit voltage (b) of a commercial thermoelectric generator for a cold side temperature of 25 $^\circ$C. Reference data is also shown. In (a) a function of the form $\eta = \alpha{}\Delta{}T^2+\delta\Delta{}T$, has been fitted to the data, while in (b) a linear function, $\Delta{}T = \gamma{}V_\n{oc}$ has been fitted.}\label{Fig.Marlow_ref}
\end{figure*}

The modules used were characterized and the efficiency as a function of temperature in air was measured using an in-house thermoelectric module tester \cite{Hung_2015}. The measured efficiency and temperature span are shown in Fig. \ref{Fig.Marlow_ref} along with datasheet reference values. For the efficiency a polynomial of order $\eta = \alpha{}\Delta{}T^2+\delta\Delta{}T$ was fitted to the data. The fit has an R$^2$ value of 0.9996, indicating a good fit to the data. Thus for the Marlow BiTe commercial module, the efficiency as a function of temperature span is given by
\begin{eqnarray}
\eta = \alpha{}\Delta{}T^2+\beta\Delta{}T
\end{eqnarray}
where $\alpha{} = -1.21(5)*10^{-6}$ K$^{-2}$ and $\beta = 4.87(7)*10^{-4}$ K$^{-1}$ for a cold side temperature of 25 $^\circ$C. Thus the efficiency for $\Delta{}T = 100$ $^\circ$C is 3.7 \% while it is 4.9 \% at $\Delta{}T = 200$ $^\circ$C.

For the temperature span as a function of open circuit voltage a linear function $\Delta{}T = \gamma{}V_\n{oc}$ was fitted to the data. Here the coefficient is $\gamma = 19.09(1)$ K V$^{-1}$.

\section{Experimental results}
In this part of the paper we consider the power produced by the constructed TEG equipped heat exchanger. Two different sets of experiments were conducted. In one, the thermal interface material between the TEGs and the pipes was varied and in the other the hot heat transfer fluid was varied. In each of these experiments the flow rate and hot inlet temperature were varied independently and the power produced by the TEGs were recorded. For each flow rate and hot inlet temperature, an I-V curve with at least 5 measurement points for all TEGs connected electrically in series were recorded, so that the maximum power could be interpolated. For selected cases, I-V curves were also made for the individual TEGs, with the remainder of the TEGs left at open circuit voltage. A wait time of 30 seconds was applied after the external load resistance was connected in order for the system to equilibrate. Measurements of voltage, current and temperature of the TEGs showed the equilibrium time to be less than $20$ seconds for all cases. In all experiments all TEGs produced almost the same amount of power, i.e. no reduction of power production was observed downstream of the hot inlet. This is due to the counterflow configuration of the experimental setup.

\subsection{Heat transfer fluids}
Two different hot heat transfer fluids were tested. These were the Evans\copyright  Power Cool 180$^\circ$ fluid and the Julabo\copyright  H20S oil. The properties of both liquids are known to the authors but are under non-disclosure agreements and cannot be reported here. Both are commercially available products. The Evans fluid has a maximum operating temperature of 180 $^\circ$C, but was for safety reason not heated above 150 $^\circ$C. The H20S oil has a maximum operating temperature of 230 $^\circ$C, but was not heated above 200 $^\circ$C. For the experiments described below, the Reynolds number for the two different fluids is between $5-100$, clearly indicating laminar flow.

A thermal paste (Omega OT-201-2), which more specifically is a silicone grease with a thermal conductivity of 2.3 W m$^{-1}$ K$^{-1}$, was used as contact material between all TEGs and pipes for this experiment. A thin layer was applied to both the TEG and pipe. A good thermal contact was ensured, as will be discussed subsequently. A total of six TEGs were used in this experimental setup. For this set of experiments, a thermocouple was inserted in the thermal paste between the TEG and the pipe, on both the hot and cold side of the TEG, for all TEGs.

\begin{figure*}[!t]
\centering
\subfigure[Evans fluid]{\includegraphics[width=1\columnwidth]{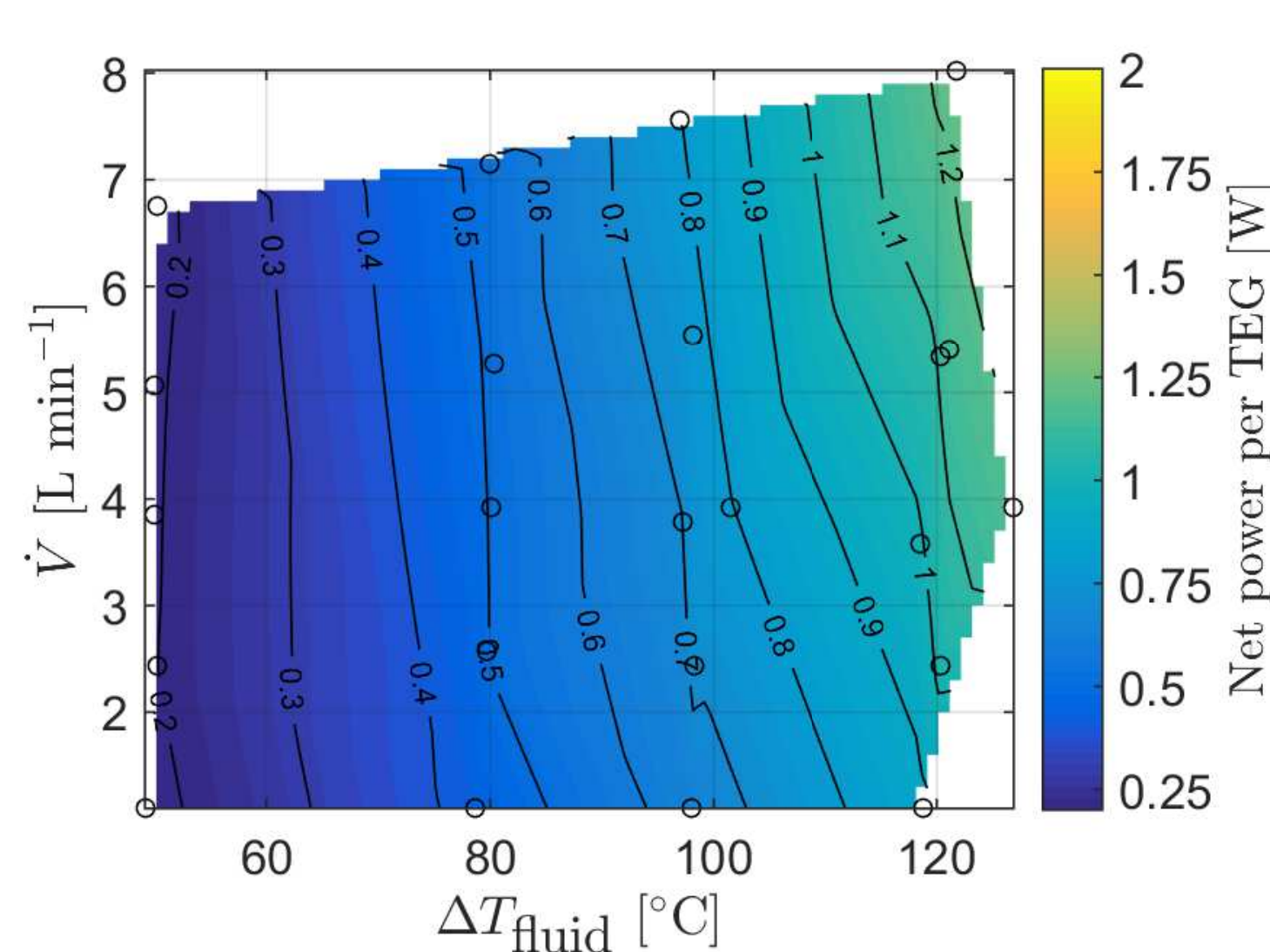}}\hspace{0.2cm}
\subfigure[H20S oil]{\includegraphics[width=1\columnwidth]{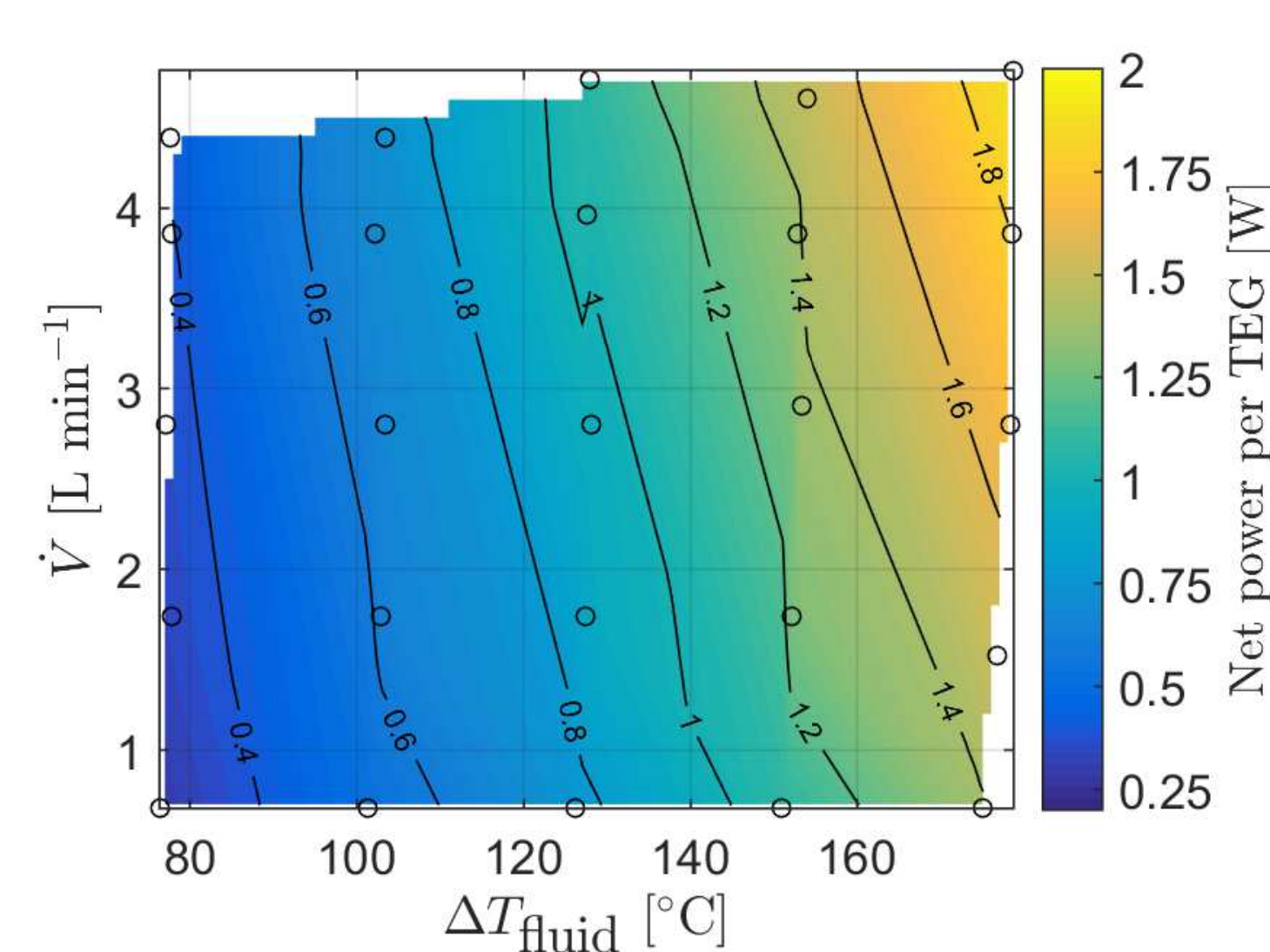}}\hspace{0.2cm}
\caption{The power per TEG as a function of the temperature difference between the inlet of the hot and cold channels and the flow rate for the Evans fluid and the H20S oil. The data points are indicated as circles and the color map is made using linear interpolation. Please note the different axes on the two figures.}\label{Fig.P_fluids}
\end{figure*}

The power generating performance of the heat exchanger as a function of the difference in hot and cold inlet temperature and the volumetric flow rate is shown in Fig. \ref{Fig.P_fluids} for the two heat transfer fluids tested. Note that the pumping power per TEG has been subtracted, thus it is the net power produced that is shown. The Evans fluid was not operated at the same maximum hot inlet temperature as the H20S oil, and thus the maximum generated power was lower, as power increases with increased temperature span. However, for the range in temperature span where the two fluids overlap, the Evans fluid is seen to outperform the H20S oil. This is due to a higher thermal conductivity of the Evans fluid as compared to the H20S oil. The pumping work of both fluids are almost identical, so it is only the thermal conductivity to the TEGs that influence the power produced. As can be seen from the Fig. \ref{Fig.P_fluids}b, a power close to 2 W per TEG was produced at the highest temperature span.

\begin{figure}[!b]
  \centering
  \includegraphics[width=1\columnwidth]{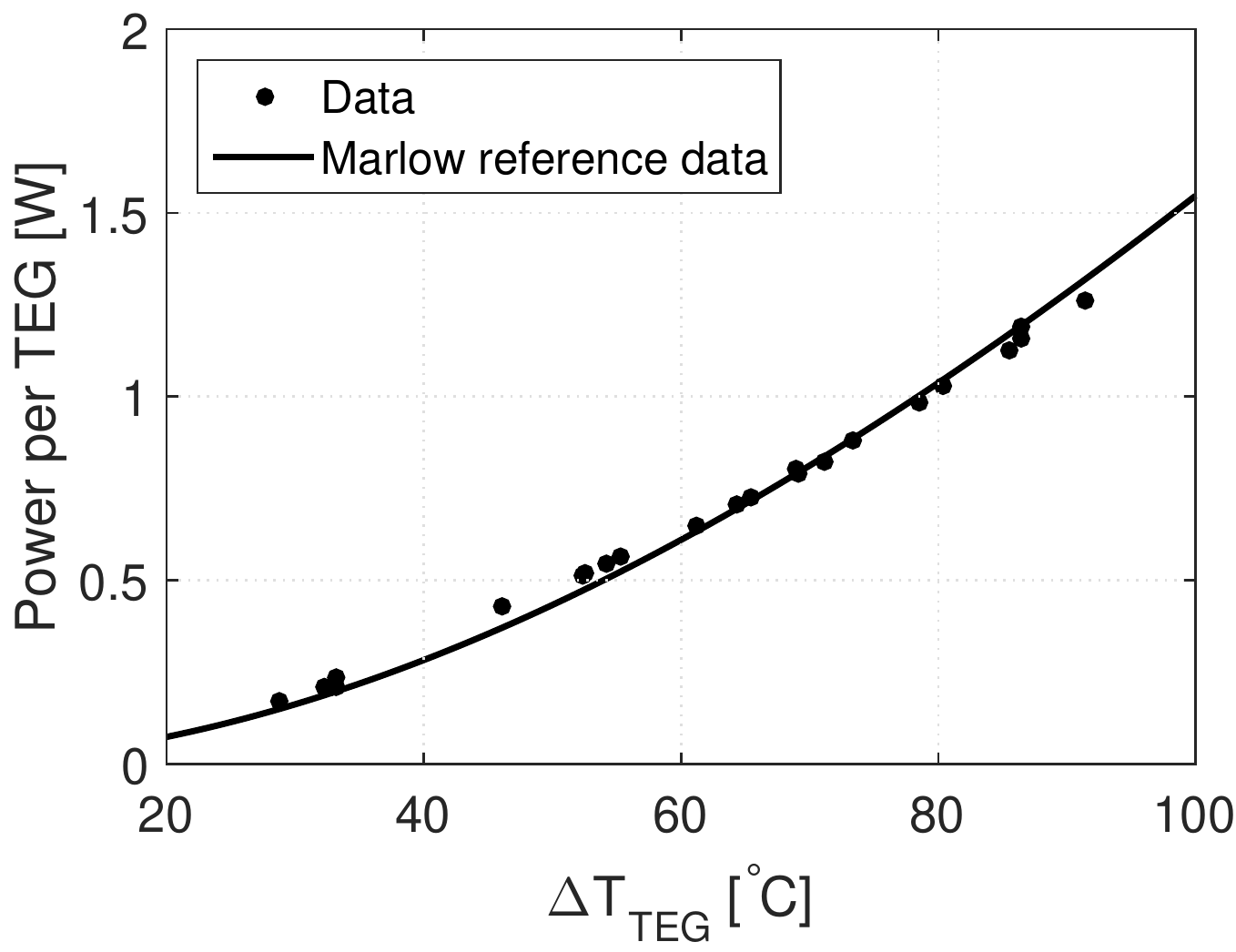}
  \caption{The measured maximum power per TEG as a function of the measured temperature span across the TEGs. Reference data is also shown.}
  \label{Fig_Power_Marlow_compare}
\end{figure}

As mentioned previously, the temperature was measured across the TEGs using thermocouples. The power produced per TEG as a function of the recorded temperature span is shown in Fig. \ref{Fig_Power_Marlow_compare} for the Evans fluid and compared with reference data. As can be seen from the figure, there is an excellent agreement between the power produced at a given temperature span and the reference data. Note that the maximum temperature span across the TEGs is 90 K, whereas the difference in temperature of the hot and cold fluids is 120 K. This is discussed in further detail below.

\subsection{Thermal interface material}
Three different interface materials between the TEGs and the pipes were tested, for selected cases. The first of these were graphite paper (Carbo Tech Nordic ApS), with a thickness of 0.2 mm, a carbon content of minimum 98 \% and an in-plane and out-of-plane thermal conductivity of 140 W m$^{-1}$ K$^{-1}$ and 5 W m$^{-1}$ K$^{-1}$, respectively. The compressibility (ASTM F36/A) is 33.55 \%. Secondly, the previously mentioned thermal paste (Omega OT-201-2) was used and finally no contact material at all between the TEG and the pipe. For this set of experiments, four TEGs were used, and the heat transfer fluid was the H20S oil. Previously, thermal interface materials that could enhance the heat flow to and from the TEG have been investigated in the medium temperature range \cite{Sakamoto_2014}, but the Omega OT-201-2 used here is not applicable in this temperature range, as it cannot be used above 200 $^\circ$C.

\begin{figure}[!b]
  \centering
  \includegraphics[width=1\columnwidth]{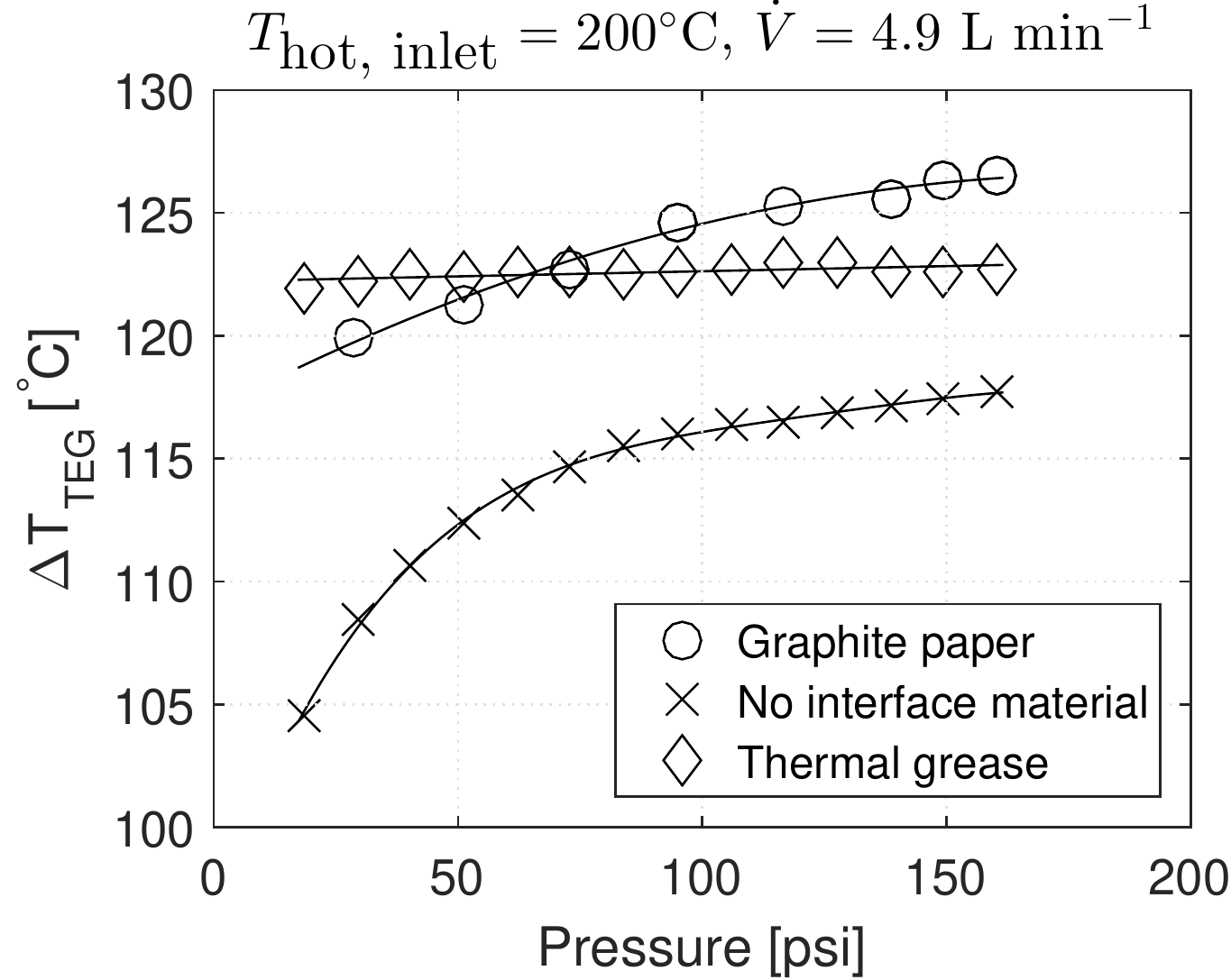}
  \caption{The temperature span as calculated from the open circuit voltage, as a function of the pressure applied on the system. The lines are guides to the eye.}
  \label{Fig_T_span_spring_compression}
\end{figure}

Initially, an experiment to determine the pressure needed to ensure a good thermal contact between the TEGs and the pipes were performed. The springs shown in Fig. \ref{Fig_Img_for_article} were gradually tightened while the open circuit voltage of the TEGs was recorded. This was done at the maximum hot inlet temperature, 200 $^\circ$C, and at the maximum flow rate $\dot{V}=4.9$ L min$^{-1}$. The temperature span across the TEGs was then calculated directly from the open circuit voltage, i.e. Fig. \ref{Fig.Marlow_ref}b, so there were no thermocouples present in this experiment. The resulting temperature span as a function of the pressure is shown in Fig. \ref{Fig_T_span_spring_compression}. The pressure is calculated by measuring the compression of the springs, the spring constant and the cross-sectional area of the TEGs.

\begin{figure*}[!t]
\centering
\subfigure[Graphite paper]{\includegraphics[width=1\columnwidth]{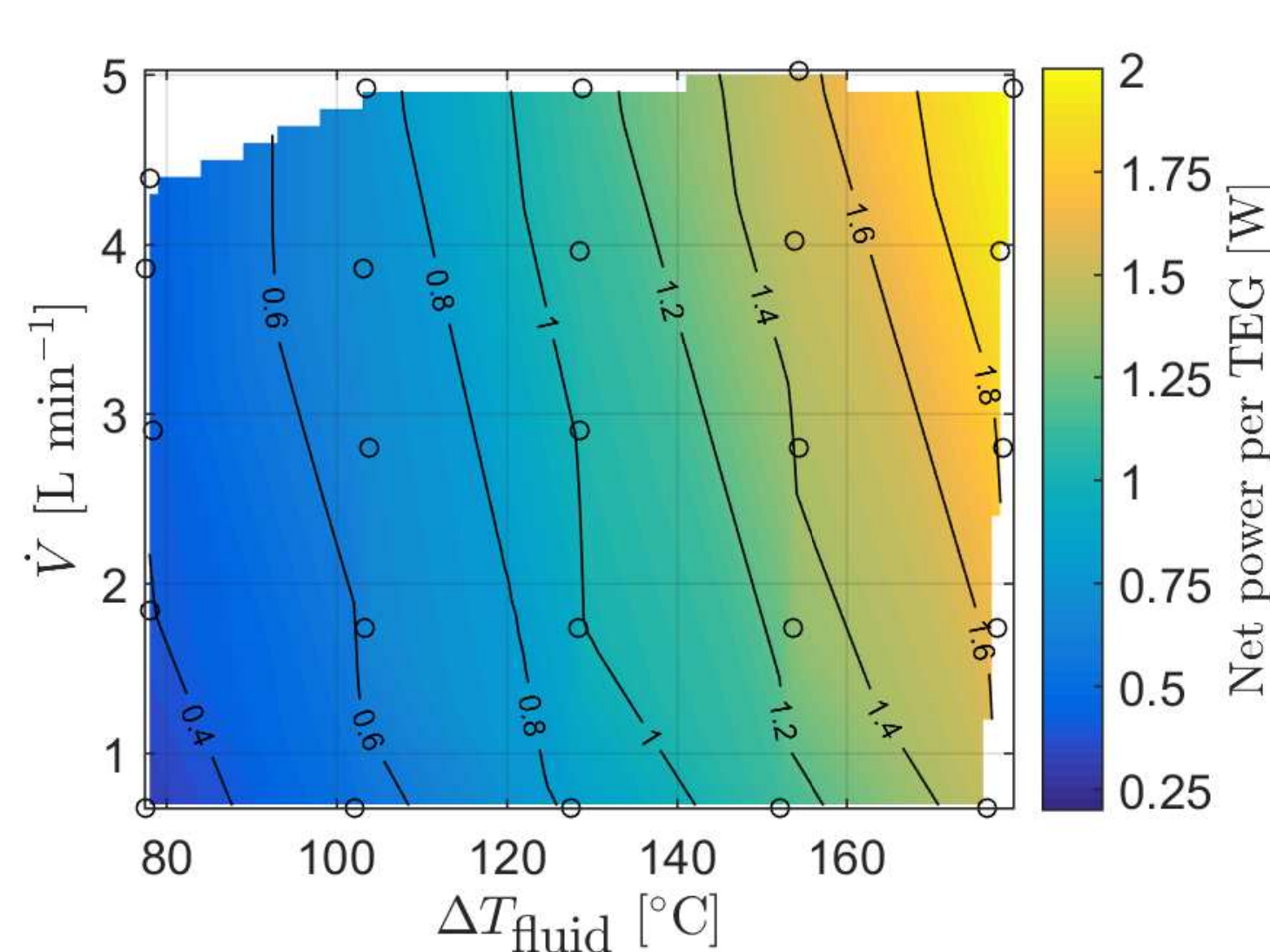}}\hspace{0.2cm}
\subfigure[No interface]{\includegraphics[width=1\columnwidth]{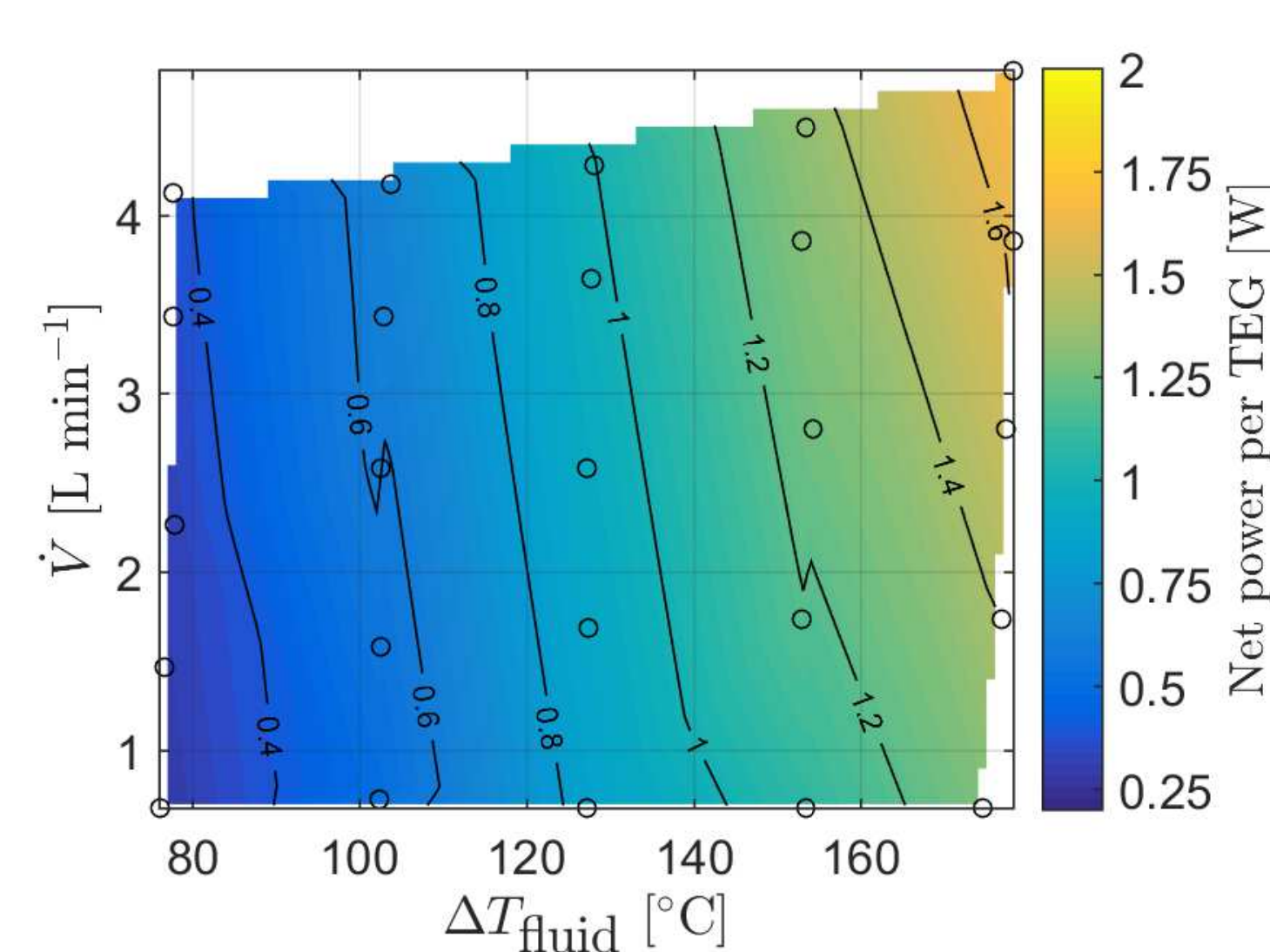}}\hspace{0.2cm}
\caption{The net power per TEG as a function of the temperature difference between the inlet of the hot and cold channels and the flow rate for graphite paper and no interface material, respectively. The pumping power per TEG has been subtracted from the power produced. The case for thermal paste as interface material is shown in Fig. \ref{Fig.P_fluids} (b). The data points are indicated as circles and the color map is made using linear interpolation.}\label{Fig.P_interface}
\end{figure*}

As can be seen from the figure, the temperature span increases as the pressure is increased. This is expected, as the surfaces are pressed more together, resulting in increased thermal conductance between the two materials. However, for the thermal paste, the temperature span across the TEGs remain constant regardless of pressure. This is because the paste fills the entire contact area even at low applied pressures and thus increasing the pressure does not improve the thermal contact. The case of no interface material is clearly seen to be the worse case tested. For the graphite paper the temperature span across the TEGs increases to above that produced by the thermal paste. This is a consequence of the higher thermal conductivity of the graphite paper as compared to that of the thermal paste. The recommended mounting pressure from the manufacturer of the TEGs is 200 psi, but it is seen from Fig. \ref{Fig_T_span_spring_compression} that a lower maximum compression pressure is needed to achieve a stable temperature span for the thermal paste. However, this is not the case for the other materials used at the interface. For the graphite paper a small increase in temperature is possible at higher pressures than those tested here, while for the case of no material at the interface a significant increase in temperature span is possible with increasing pressure.

Having determined the temperature span across the TEGs as a function of the applied pressure, the net power produced per TEG was measured as a function of flow rate and temperature difference at the highest pressure of 160 psi. Note that again the pumping power per TEG has been subtracted. The results are shown in Fig. \ref{Fig.P_interface} for the graphite paper and the case where no material was placed at the interface, while the case for the thermal paste is shown in Fig. \ref{Fig.P_fluids}b. The power produced is seen to be highest in the case of graphite paper where a maximum power of 2 W per TEG or 0.22 W cm$^{-2}$ was produced at a hot inlet temperature of 200 $^\circ$C and a flow rate of 4.9 L min$^{-1}$.

\begin{figure*}[!t]
\centering
\subfigure[Specific $\dot{V}$]{\includegraphics[width=1\columnwidth]{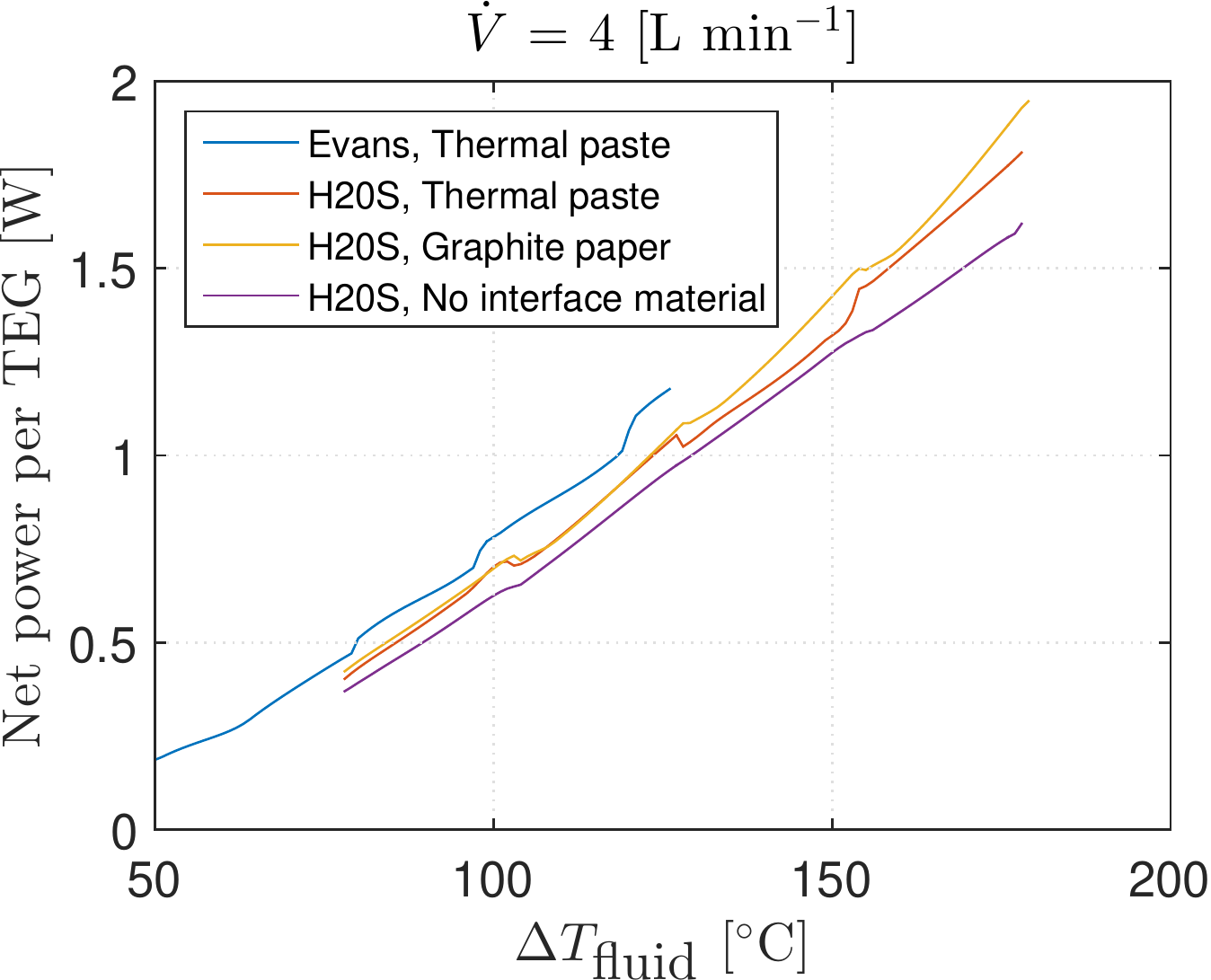}}\hspace{0.2cm}
\subfigure[Specific $\Delta{}T_\n{fluid}$]{\includegraphics[width=1\columnwidth]{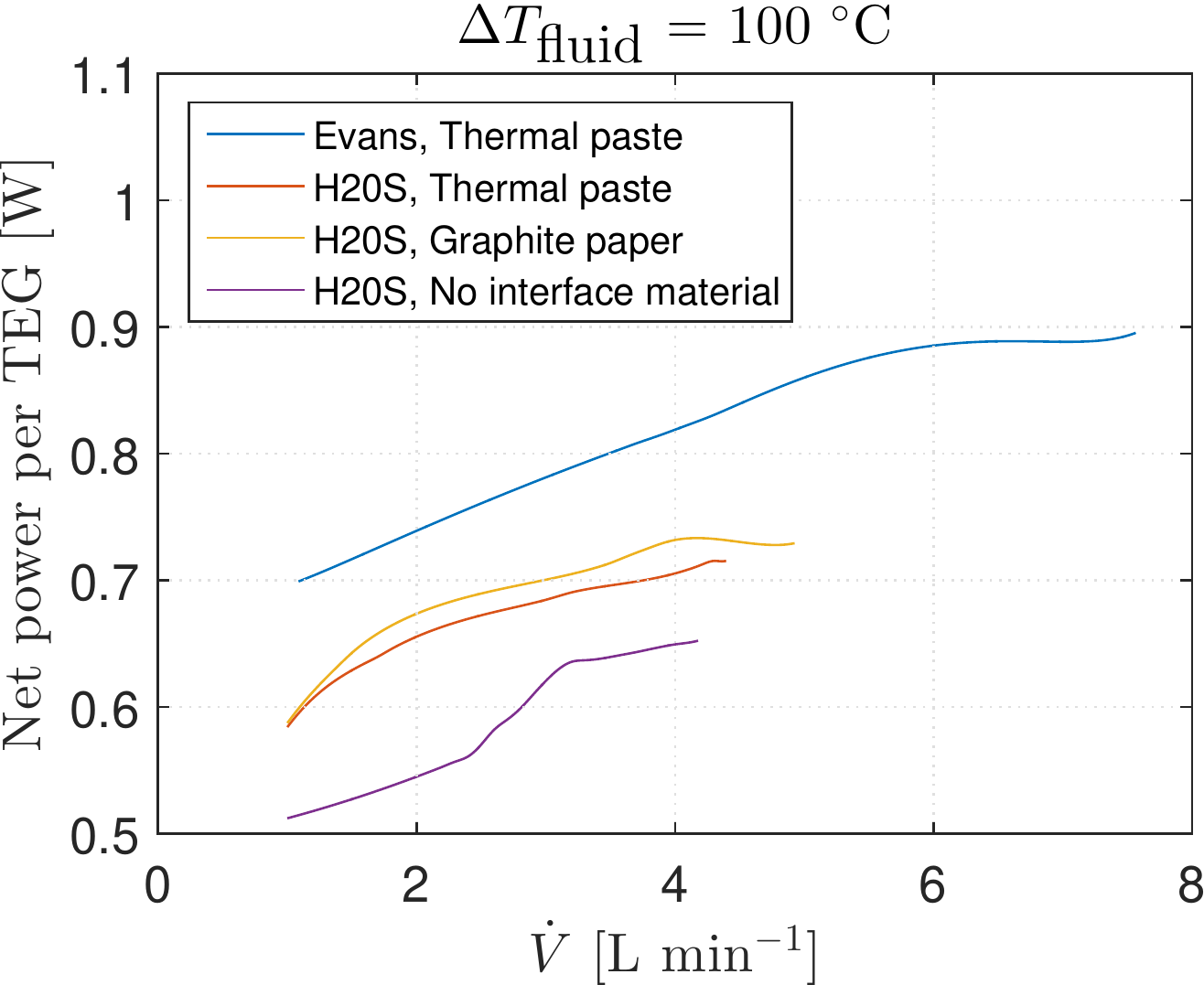}}\hspace{0.2cm}
\caption{The power per TEG for all considered system for (a) a specific flow rate, $\dot{V}=4$ L min$^{-1}$, and (b) for a specific fluid temperature difference, $\Delta{}T_\n{fluid}=100$ K. The data has been interpolated using spline interpolation from the data shown in Figs. \ref{Fig.P_fluids} and \ref{Fig.P_interface}.}\label{Fig.P_compare}
\end{figure*}

In order to better compare the different heat transfer fluids and thermal interface materials, the power per TEG is shown for a specific flow rate and for a specific temperature difference in Fig. \ref{Fig.P_compare} for all cases considered. As can be seen the Evans fluid, in the same temperature range, produces a higher power per TEG than the H20S oil. The power per TEG as a function of flow rate is seen to flatten out with increased flow rate, as also observed in numerical simulations \citep{Henderson_1979}. A relative increase in the inlet temperature is seen to result in a much greater increase in power production than a relative increase in the flow rate, as also observed by \citet{Esarte_2001}.

\subsection{Temperature span across the TEGs}
\begin{figure}[!t]
  \centering
  \includegraphics[width=1\columnwidth]{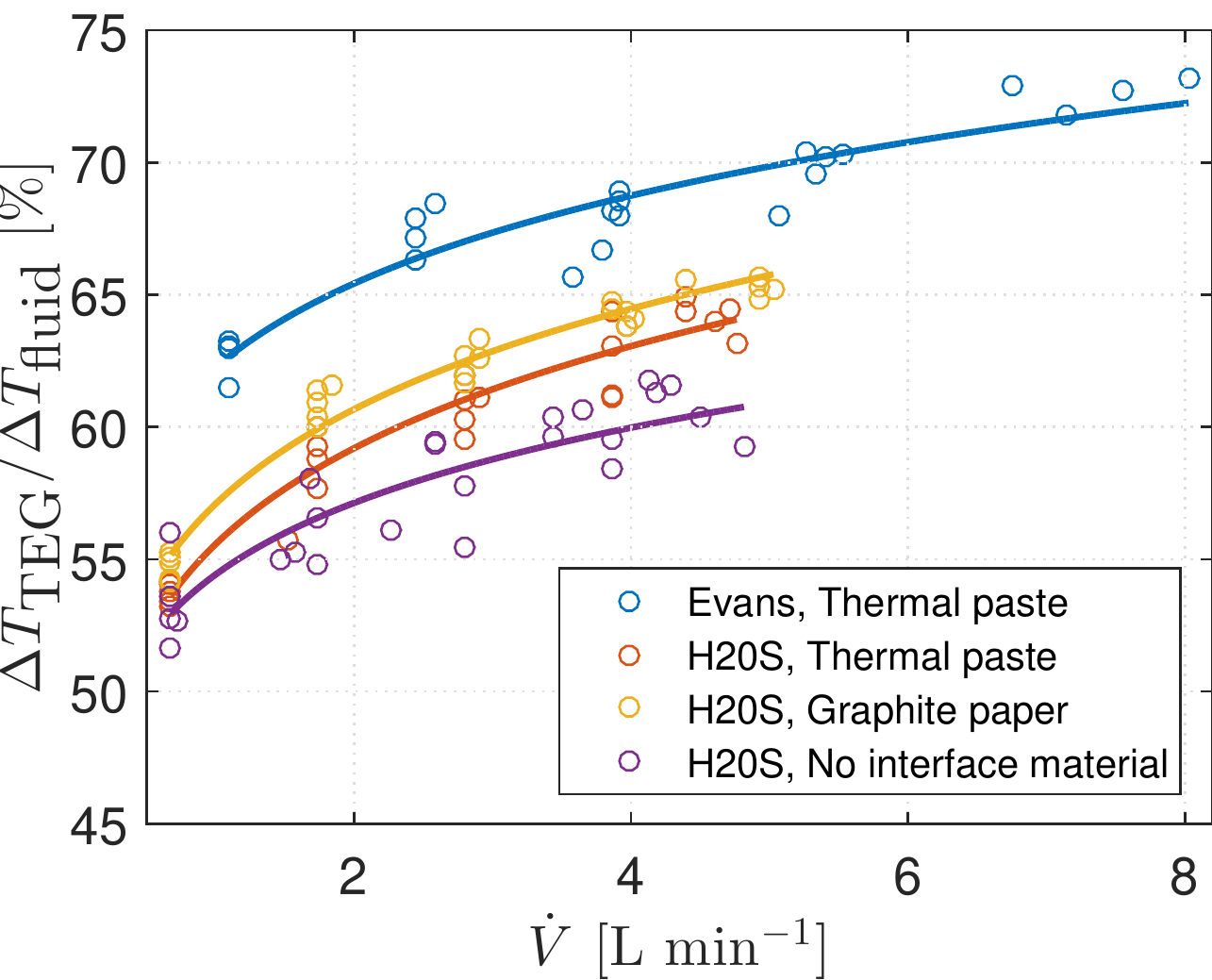}
  \caption{The difference in percentage between the temperature difference in the fluid and the temperature difference across the TEGs. The lines are a guide to the eye. A temperature span across the TEGs of 55\% to 75\% of that of the fluid is realized.}
  \label{Fig_T_diff_flow_DT}
\end{figure}

In order to investigate the actual temperature span across the TEGs realized in the experimental setup, this has been calculated for the different fluids and thermal interface materials, based on the produced power, as shown in Fig. \ref{Fig_Power_Marlow_compare}. Normalizing the temperature difference to the temperature difference of the fluid, the result become a function of the flow rate and not the absolute temperature. This is shown in Fig. \ref{Fig_T_diff_flow_DT} where the TEG temperature difference in percent is shown as function of flow rate. The realized temperature span across the TEGs is seen to be 55\% to 75\% of the temperature span across the fluid. Thus, there is room for improving the heat exchanger, which would lead to a high power generating system. However, the realized temperature spans are still an improvement over the value of 40\% reported by \citet{Crane_2004} for a similar experimental setup. As the power produced by the TEG is not linear in the temperature span, as shown in Fig. \ref{Fig_Power_Marlow_compare}, could the last 25-45\% of the temperature difference be realized, the power produced per TEG would increase to 2.2 W per TEG for the Evans fluid at 150 $^\circ$C and 3.8 W per TEG for the H20S oil at 200 $^\circ$C, both of which are substantial increases in power production.

\section{A scaled-up experiment}
\begin{figure}[!t]
  \centering
  \includegraphics[width=1\columnwidth]{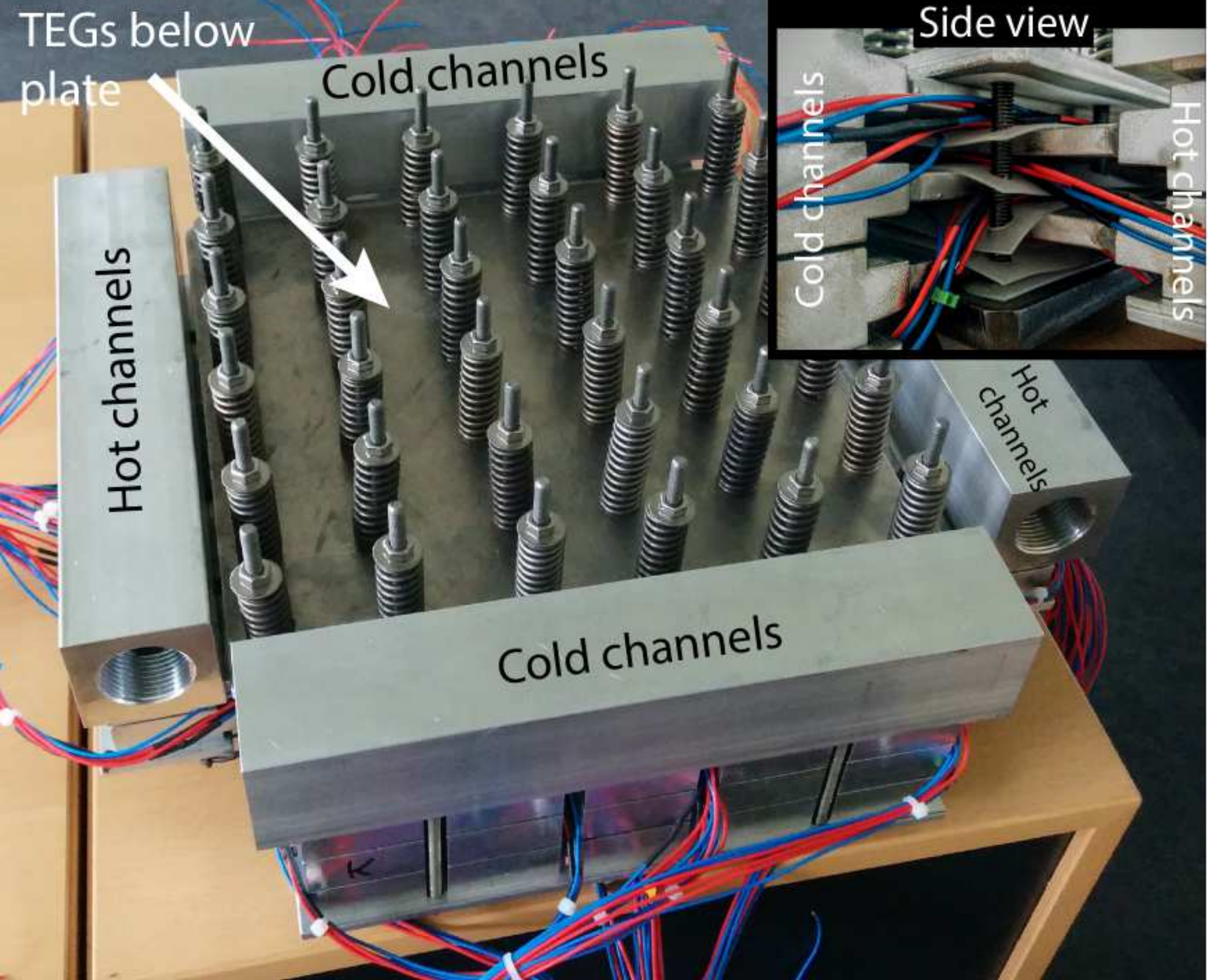}
  \caption{An image of the large scale system, before installation in the experimental setup. The entire device has an area of 40x40 cm$^2$, while a single fluid channel is 30 cm long. The TEGs are located below the top plate. The side view shows the matrix orientation of the hot and cold flow channels.}
  \label{Fig_Img_for_article_large_scale}
\end{figure}

A much larger version of the experimental setup described above was constructed and tested, to verify power production at a larger scale. A setup with 5x2 hot fluid channels and 5x3 cold fluid channels, arranged in a matrix, and with a total of 100 TEGs at the intersections between the hot and cold channels were constructed. This resulted in four ``layers'' of TEGs, with 25 TEGs in each layer. Graphite paper were used as the thermal interface material. A photograph of the experimental setup is shown in Fig. \ref{Fig_Img_for_article_large_scale}.

\begin{figure}[!t]
  \centering
  \includegraphics[width=1\columnwidth]{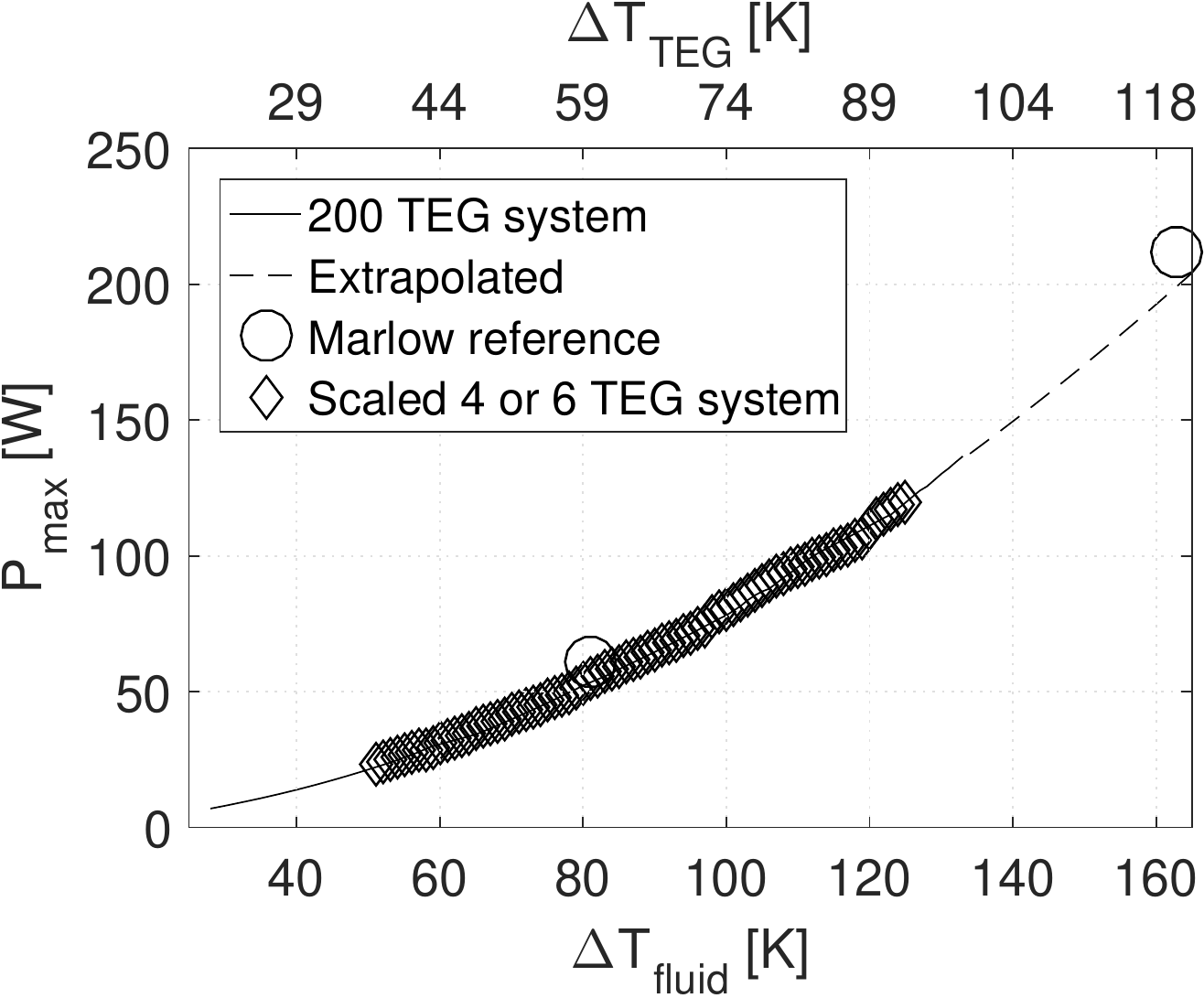}
  \caption{The maximum power as function of temperature span of the fluid (bottom $x$-axis) or across the TEGs (top $x$-axis). The temperature span across the TEGs is interpolated based on TEG properties and the produced power, while the temperature span of the fluid is measured. Reference data (Marlow), as well as data interpolated from the small scale experiment is shown.}
  \label{Fig_Max_power_vs_temp}
\end{figure}

The system was tested similarly to the small scale experiment described above, albeit only at the maximum possible hot flow rate of 9 L min$^{-1}$ per channel and only using the Evans liquid, for simplicity. The produced open circuit voltage as function of both hot fluid temperature and applied pressure was measured, as well as I-V curves at an applied pressure of 48 psi at different hot fluid temperatures. Note that this applied pressure is lower than the previously used 150 psi. This was due to availability of the used springs. The flow rate in the large scale experiment was higher than that in the small scale experiment. The increase in temperature span by increasing the pressure from 50 psi to 150 psi in the small scale experiment can be extrapolated from Fig. \ref{Fig_T_span_spring_compression}, while the decrease in temperature span due to the decrease in flow rate can be extrapolated from Fig. \ref{Fig_T_diff_flow_DT}. Performing the calculations, it can be shown that scaling the small system from 150 psi to 50 psi results in a temperature drop across the TEGs that is the same as if the flow rate was changed from 9 L min$^{-1}$ to 5 L min$^{-1}$. Therefore, the large scale setup should be compared to the small scale experiment at this flow rate. The produced maximum power as function of temperature span is shown in Fig \ref{Fig_Max_power_vs_temp}, along with reference data and scaled-up results of the small experimental setup. As can be seen from the figure, the small and large scale experiments produce the same power per TEG. This shows that power production on a larger scale using waste heat is possible.

\section{Discussion and perspectives}
An experimental realization of a heat exchanger producing electricity is an important step towards applications using TEGs. However, such a device cannot be optimized purely using experimental techniques, as this would be much too cumbersome, time-consuming and costly. Therefore, a numerical model has been developed that can be used to dimension the optimal heat exchanger geometry. This is discussed in a subsequent work \citep{Sarhadi_2016}.

Regarding the power produced by the experimental setup discussed above, the results are comparable to the 1.8 W per TEG found by \citet{Yu_2007} using a numerical model, albeit with TEGs with a quarter of the area considered here. However, their heat transfer fluid was pressurized water, which has a thermal conductivity about a factor of five higher than the fluids considered here, which can easily account for the increased power production. However, the power produced by the system realized here is only about half of that of the system by \citet{Niu_2009}. This is caused by a limited heat transfer from the fluid to the TEGs. This can be seen in Fig. \ref{Fig_T_span_spring_compression}, where  a maximum temperature span across the TEGs of $\Delta{}T=175$ K should be expected for the H20S oil in the case of perfect heat transfer from the fluid to the TEGs. However, only a temperature difference of $\Delta{}T=125$ K is realized. This is not the case for the system by \citet{Niu_2009}, which realizes the full temperature span of the fluids across the TEGs, leading to a significant increase in power production.

It is also of interest to discuss the power generation efficiency of the heat exchanger. As the pumping work is negligible at high temperature, the efficiency of the system is given solely by the thermoelectric conversion efficiency of the TEGs in the system, as shown in Fig. \ref{Fig.Marlow_ref}. From Fig. \ref{Fig_T_span_spring_compression} it is seen that the maximum realized temperature span across the TEGs is close to $\Delta{}T_\n{TEG}=125$ $^\circ$C. The corresponding efficiency is calculated to be 4.1\%, directly from Fig. \ref{Fig.Marlow_ref}, assuming no power has to be spend on cooling the cold liquid. The power conversion efficiency is inherently limited by the efficiency of the TEGs. If the system could be equipped with TEGs with a higher conversion efficiency, the increase in efficiency of the system would increase proportionally.

\section{Conclusion}
An experimental realization of a heat exchanger with TEGs was presented and discussed in detail. The power producing capabilities as a function of flow rate and temperature span were characterized for two different commercial heat transfer fluids, and for three different thermal interface materials. The device is shown to be able to produce 2 W per TEG or 0.22 W cm$^{-2}$ at a fluid temperature difference of 175 $^\circ$C and a flow rate per channel of 5 L min$^{-1}$, for both a small scale device with 4-6 TEGs and a large scale device with 100 TEGs. The optimal thermal interface material was shown to be graphite paper. Furthermore, the power production was shown to depend more critically on the fluid temperature span than on the fluid flow rate. Finally, the temperature span across the TEG was shown to be 55\% to 75\% of the temperature span of the hot and cold fluids.

\section*{Acknowledgements}
The authors would like to thank The Energy Technology Development and Demonstration Program (EUDP), Danish Energy Agency for sponsoring this research project as Project No. 64012-0155. The authors would also like to acknowledge Dansk Teknologi for assisting with the construction of the large scale prototype.

\end{document}